# Does Life Come from Water or Fluids?

Piero Chiarelli

*National Council of Research of Italy, Area of Pisa, 56124 Pisa, Moruzzi 1, Italy Interdepartmental Center*
*"E.Piaggio" University of Pisa*
Phone: +39-050-315-2359
Fax: +39-050-315-2166

Email: pchiare@ifc.cnr.it.

In the present work the Stochastic generalization of the quantum hydrodynamic analogy (SQHA) is used to obtain the far from equilibrium kinetics for a real gas and its fluid phase.
In gasses and their liquids, interacting by Lennard-Jones potentials whose mean distance is bigger than the quantum correlation distance and than the molecular interaction distance $r_0$, it is possible to define a Fokker-Plank type equation of motion as a function of the mean phase space molecular volume that far from equilibrium shows maximizing the dissipation of a part of the generalized SQHA-free energy.
In the case of a real gas with no chemical reactions and at quasi-isothermal conditions, the principle disembogues into the maximum free energy dissipation confirming the experimental outputs of electro-convective instability.
In this case, the model shows that the transition to stationary states with higher free energy can happen and that, in incompressible fluids, the increase of free energy is almost given by a decrease of entropy leading to the appearance of self-ordered structures.
The output of the theory showing that the generation of order, via energy dissipation, is more efficient in fluids than in gasses, because of their incompressibility, leads to the re-conciliation between physics and biology furnishing the eplanation why the life was born in water.
The theoretical output also suggests that the search for life out of the earth must consider the possibility to find it in presence of liquid phases different from water.

## INTRODUCTION

The missing bridge between the physics and life disciplines constitutes one of the greatest problem of nowadays science. The situation is similar to that existing between physics and chemistry at the beginning of the past century. Every scientist was intimately convinced that once the physical law of particles belonging to the atoms were established the chemical properties of elements should possibly derived and explained.
With the advent of quantum mechanics this view came to realization and the quantum physics is at the base of modern chemistry.
A similar state of fact exists between the physics and biology. The unavailability of a physical theory capable of explaining the order generation at the base of the evolution of biological system is the major obstacle to give the physical basis to the biology.
One of the most known and experimentally verified law of the physics is the second law of thermodynamic that expresses the tendency of any physical system toward the maximum possible disorder. Even verified only in the near-equilibrium condition, this law has lead many thinkers to a "pessimistic" view of the world and to the so called "obscurantism" in many field of human science.
Starting by the thirties of the last century, mainly due to the work of Prigogine, it was born in the physicists and chemists the idea that the tendency to the entropy production was a bounded process and that generation of order were possible out of the neighborhoods of thermodynamic equilibrium [1-8]. Various principles have been proposed for the self-organized régimes governed by classical linear and non-linear non-equilibrium thermodynamic laws, with stable stationary configurations being particularly investigated.
Nevertheless an organic understanding for a long time has been unavailable. In 1945 Prigogine [1,2] proposed the "Theorem of Minimum Entropy Production" which applies only to near-equilibrium stationary state. The proof offered by Prigogine is open to serious criticism [3]. Šilhavý [4] offers the opinion that the extremal principle of [near-equilibrium] thermodynamics does not have any counterpart in far from-equilibrium steady states despite many claims in the literature.

Sawada postulated the principle of largest amount of entropy production [5]. He started by the work given by Malkus and Veronis [6] about the earth's atmospheric turbulence where the principle of maximum heat current, holding in fluid mechanics, has been proven to drive the energy transport process. Sawada showed that for a given boundary condition this law corresponds to the maximum entropy production, but this inference is not ever valid.

Sawada and Suzuky showed that the maximum entropy production leads in electro-convective phenomena to the maximum rate of energy dissipation. This principle was confirmed, both by numerical simulations and by experiments [8], in electro-convective instabilities. Moreover, they showed that ordered metastable ordered states were visited by the system with a living time proportional to the rate of energy dissipation .

The rate of dissipation of energy appeared for the first time in Onsager's work [7] on this subject. An extensive discussion of the possible principles of maximum entropy production and/or of dissipation of energy is given by Grandy [9]. He finds difficulty in defining the rate of internal entropy production in the general case, showing that sometimes, for the prediction of the course of a process, the maximum rate of dissipation of energy may be more useful than that of the rate of entropy production.

Nowadays, the debate about the principle of maximum free energy dissipation (MFED) and the Prigogine one's is come to a possible solution. The author has shown that in the frame of the quantum stochastic approach of hydrodynamic analogy [10-13], it is possible to define a phase space Wigner-type distribution functions that in (classical) phases, were the mean inter-particle distance is  bigger than the distance over which the quantum correlations take place, can lead to the definition of an energy function (named stochastic free energy) whose dissipation is maximum during the out of equilibrium relaxation process.

Following the tendency to reach the highest rate stochastic free energy dissipation (SFED), a system relaxing to equilibrium goes through states with higher order so that the matter self-organization becomes possible.

Moreover, near equilibrium the maximum of SFED is shown to lead to the Prigogine's principle of minimum entropy production, while far from equilibrium, in quasi-isothermal states and in the case of elastic molecular collisions and in absence of chemical reactions, the maximum SFED reduces to the maximum free energy dissipation.

The Prigogine's principle of minimum entropy production near-equilibrium and the far from equilibrium Sawada's principle of maximum energy dissipation are two complementary principia of a more general theory that do not contradict each other. The availability of a clear general model  applying to the irreversible processes far from equilibrium is a great tool to investigate the generation of order and matter self-assembling in presence of large gradients of thermodynamic forces.

In the present paper the author shows that in quasi-isothermal far from equilibrium states with no chemical reactions, the principle of maximum SFED, that can generally lead to transition to a state with an increase of free energy, in the case of incompressible phase (e.g., fluid one) this increase of free energy is basically almost given by an increase of order since the energy variation of the system associated to the external work is quite null.

For the first time it comes out from a physical theory the information about how and why a fluid phase is necessary to the development of matter self-assembling process that is at the base of the emerging of life.

The physical model is not devoid of new information signaling that ordering processes (and life) can also happen in fluids different from water, an interesting chance in searching for life in the universe out of the earth.

## 2. The SQHA equation of motion

The quantum hydrodynamic analogy (QHA) equations are based on the fact that the Schrödinger equation, applied to a wave function $\psi_{(q,t)} = A_{(q,t)} exp[i\frac{S_{(q,t)}}{\hbar}]$, is equivalent to the motion of a fluid owing the particle density $n_{(q,t)} = A^2_{(q,t)} = |\psi|^2$ with a velocity $\dot{q} = \frac{\nabla_q S_{(q,t)}}{m}$, governed by the equations [10-14]

$$\partial_t n_{(q,t)} + \nabla_q \cdot (n_{(q,t)} \dot{q}) = 0, \quad (1)$$

$$\dot{q} = \nabla_p H, \quad (2)$$

$$\dot{p} = -\nabla_q (H + V_{qu}), \quad (3)$$

where $\nabla_p = \left(\frac{\partial}{\partial p_i}\right)$, $\nabla_q = \left(\frac{\partial}{\partial q_i}\right)$, $H$ is the Hamiltonian of the system and $V_{qu}$ is the quantum pseudo-potential that reads

$$V_{qu} = -\left(\frac{\hbar^2}{2m}\right)n^{-1/2}\nabla_q \bullet \nabla_q n^{1/2}. \qquad (4)$$

For the phase space analysis, it is useful to observe that equations (1-3) can be derived by the following phase-space equations [13]

$$\partial_t \rho_{(q,p,t)} + \nabla \bullet (\rho_{(q,p,t)}(\dot{x}_H + \dot{x}_{qu})) = 0 \qquad (5)$$

where $\nabla = \left(\frac{\partial}{\partial q_i}, \frac{\partial}{\partial p_i}\right)$ and where

$$n_{(q,t)} = \iiint \rho_{(q,p,t)} d^{3n}p. \qquad (6)$$

$$\dot{x}_H = (\nabla_p H, -\nabla_q H) \qquad (7)$$

$$\dot{x}_{qu} = (0, -\nabla_q V_{qu}) \qquad (8)$$

where

$$\rho_{(q,p,t)} = n_{(q,t)}\delta(p - \nabla_q S) \quad , \qquad (9)$$

and where

$$S = \int_{t_0}^{t} dt\left(\frac{p \bullet p}{2m} - V_{(q)} - V_{qu}\right), \qquad (10)$$

The Madelung approach, as well as the Schrödinger one, are non-local and are not able to give rise to local limit. When fluctuations are added to the QHA equation of motion, the resulting stochastic-QHA (SQHA) dynamics shows that is possible to obtain a local dynamics on large scale, preserving the quantum behavior on a microscopic one. In a preceding paper [13] the author has shown that in presence of vanishing small stochastic Gaussian noise, the QHA motion equation (at first order of approximation in the noise amplitude $\Theta$) reads

$$\partial_t \rho_{(q,p,t)} = -\nabla \bullet (\rho_{(q,p,t)}(\dot{x}_H + \dot{x}_{qu(n)})) + \eta_{(q,t,\Theta)}\delta(p - \nabla_q S), \qquad (11)$$

where the phase space distribution

$$\rho_{(q,p,t)} = n_{(q,t)}\delta(p - \nabla_q S) \qquad (12)$$

is a Wigner-like distribution obeying to the property in the limit of null noise the property

$$|\psi|^2 = \int_{-\infty}^{+\infty} d^{3n}p \rho_{(q,p,t)}, \qquad (13)$$

where $\Theta$ is a measure of the vacuum noise amplitude (VNA) and

$$<\eta_{(q_\alpha,t)}, \eta_{(q_\beta+\lambda,t+\tau)}> = \mu\frac{k\Theta}{\lambda_c^2}\exp\left[-\left(\frac{\lambda}{\lambda_c}\right)^2\right]\delta(\tau)\delta_{\alpha\beta} \qquad (14)$$

is the VNA variance, where the quantum correlation length $\lambda_c$ reads [13]

$$\lambda_c = \pi\frac{\hbar}{(2mk\Theta)^{1/2}} \quad , \qquad (15)$$

$$S = \int_{t_0}^{t} dt(\frac{p \bullet p}{2m} - V_{(q)} - V_{qu(n)}) = \int_{t_0}^{t} dt(\frac{p \bullet p}{2m} - V_{(q)} - V_{qu(n_0)} - I^*) \quad (16)$$

$$m\dot{q} = p = \nabla_q S = \nabla_q \{\int_{t_0}^{t} dt(\frac{p \bullet p}{2m} - V_{(q)} - V_{qu(n_0)} - I^*)\} = p_0 + \Delta p_{st}, \quad (17)$$

where

$$\Delta p_{st} = -\nabla_q \{\int_{t_0}^{t} I^* dt\}, \quad (18)$$

where

$$I^* = V_{qu(n)} - V_{qu(n_0)} \quad (19)$$

where $n_0$ is obtained from the zero order of approximation equation (1).

## 3. Macroscopic local limiting dynamics

Given $\Delta L$ the physical length of the system, the macroscopic local dynamics is achieved for those problems that satisfy the condition

$$\lambda_c \cup \lambda_q \ll \Delta L.$$

From the condition $\lambda_q \ll \Delta L$ it follows that [13]

$$\lim_{q/\lambda_q \to \infty} -\nabla_q V_{qu(n_0)} = 0 \quad (20)$$

and the SPDE of motion acquires the form [13]

$$\partial_t \rho_{(q,p,t)} = -\nabla \bullet (\rho_{(q,p,t)} \dot{x}_H) + \eta_{(q,t,\Theta)} \delta(p - \nabla_q S) \quad (21)$$

$$\rho_{(q,p,t)} = n_{(q,t)} \delta(p - \nabla_q S), \quad (22)$$

$$\partial_t n_{(q,t)} = -\nabla_q \bullet (n_{(q,t)} \dot{q}_{cl}) + \eta_{(q_\alpha,t,\Theta)} \quad (23)$$

$$<\eta_{(q_\alpha,t)}, \eta_{(q_\alpha+\lambda,t+\tau)}> = \underline{\mu} \delta_{\alpha\beta} \frac{2k\Theta}{\lambda_c} \delta(\lambda)\delta(\tau) \quad (24)$$

$$\dot{q} = \frac{p}{m} = \lim_{\Delta\Omega/\lambda_c \to \infty} \lim_{\Delta\Omega/\lambda_q \to \infty} \frac{\nabla_q S}{m}$$

$$= \nabla_q \{\lim_{\Delta\Omega/\lambda_c \to \infty} \lim_{\Delta\Omega/\lambda_q \to 0} \frac{1}{m} \int_{t_0}^{t} dt(\frac{p \bullet p}{2m} - V_{(q)} - V_{qu(n)} - I^*)\} \quad (25)$$

$$= \frac{1}{m} \nabla_q \{\int_{t_0}^{t} dt (\frac{p \bullet p}{2m} - V_{(q)} - \Delta)\} = \frac{p_{cl}}{m} + \frac{\delta p}{m} \cong \frac{p_{cl}}{m}$$

where $\delta p$ is a small fluctuation of momentum and

$$\dot{p}_{cl} = -\nabla_q V_{(q)}, \quad (26)$$

Being also $\lambda_c \ll \Delta L$, $I^*$ represents a small energy fluctuation due to the quantum potential [13].

## 4. The kinetic equation for classical gas and fluid phases

As derived in reference [14], for a gas or mean-field fluid phases, we can describe our system by a single particle SQHA distribution function (DF) $\rho_{(1)}$ from which we can extract the statistical single particle distribution $\rho^s$ that obeys to the equation.

$$\lim_{\Delta L \gg \lambda_c, \lambda_q} \partial_t \rho^s + \nabla \bullet (\rho^s (<\dot{x}_{\overline{H}}> + <\dot{x}_s>)) = 0 \qquad (27)$$

where $<\dot{x}_s>$ obeys to the equation [14]

$$\lim_{\Delta L \gg \lambda_c, \lambda_q} \rho^s <\dot{x}_s> = -\nabla D_{(i)} \rho^s \qquad (28)$$

where

$$\dot{x}_{\overline{H}} = (\frac{\partial \overline{H}}{\partial p}, -\frac{\partial \overline{H}}{\partial q}), \qquad (29)$$

where $\overline{H} = H + \overline{V}$ is the mean-field Hamiltonian
Equation (28) is basically the Fokker-Plank form of the Maxwell equation.
In order to obtain from (28) a closed kinetic equation, the standard approach is to introduce additional information about the diffusion coefficient $D$. Under the local equilibrium condition this is usually achieved by the semi-empirical assumption of linear relation between flows and fluxes.
In order to obtain an evolutionary principle far from equilibrium, here we use (28) without the assumption of linear relation between flows and thermodynamic forces.

### 4.1 The mean phase space molecular volume of WFM

In order to grasp information from (28) we observe that (for gasses and mean-field fluids) the SQHA approach shows two competitive dynamics: (a) the enlargements of the molecular DF [14] between two consecutive collisions, (b) The diffusion of the molecules, in term of their mean position, as a consequence of the molecular collisions (that cause the WFM collapse [15]).
As consequence of free expansions and collapses, the pseudo-Gaussian molecular DF in the phase space cell $\Delta\Omega$ will occupy the mean volume $<\Delta V_m>$ that we pose

$$\lim_{\Delta L \gg \lambda_c, \lambda_q} <\Delta V_m> = h^3 \exp[-\phi] \qquad (30)$$

where $<\Delta V_m>$ reads:

$$<\Delta V_m> = \frac{\sum_{i \in \Delta\Omega} (\int_{\Delta\Omega_{(q,p)}} \rho_{(i)} (x_{(i)} - <x_{(i)}>)^2 d^3q d^3p)^{\frac{1}{2}}}{\sum_{i \in \Delta\Omega} \int_{\Delta\Omega_{(q,p)}} \rho_{(i)} d^3q d^3p} \qquad (31)$$

$$<x_{(i)}> = \frac{\int_{\Delta\Omega_{(q,p)}} \rho_{(i)} x_{(i)} d^3q d^3p}{\int_{\Delta\Omega_{(q,p)}} \rho_{(i)} d^3q d^3p} \qquad (32)$$

where $x_{(i)} = \begin{pmatrix} q_{(i)} \\ p_{(i)} \end{pmatrix}$.

In the case of sufficiently weak radiative coupling between vacuum fluctuations and thermal ones [14], (28) simplifies as

$$\lim_{\Delta L \gg \lambda_c, \lambda_L} <\dot{x}_S> = -D\nabla\phi(1 + A + O(\nabla\phi^2)) \qquad (33)$$

that introduced into the FPE (27) leads to the kinetic equation

$$\partial_t \rho^S + \nabla \bullet \rho^S <\dot{x}_{\overline{H}}> = \nabla \bullet \rho^S D\nabla\phi(1 + A + O(\nabla\phi^2)) \qquad (34)$$

## 4.2 Far from equilibrium relaxation and maximum stochastic free energy dissipation in stationary states

Even if the $\phi$-function is well defined far from equilibrium, without the explicit form of the diffusion coefficient and the initial and boundary condition of an assigned problem, the kinetic equations (27-28) is just a symbolic equation.

Nevertheless, the existence of the $\phi$-function far from equilibrium allows the definition of a formal criterion of evolution.

In fact from equation (33) it is possible to derive an evolutionary principle along the relaxation pathway can be formulated in terms of dissipation of the $\phi$-function (named here *normalized hydrodynamic free energy* (NHFE) since at equilibrium it converges to the free energy normalized to kT [14].

Given that, the total differential of the *normalized hydrodynamic free-energy* $\phi$ can be written as a sum of two terms, such as:

$$\frac{d\phi}{dt} = \frac{\partial\phi}{\partial t} + <\dot{x}> \bullet \nabla\phi = \frac{\partial\phi}{\partial t} + <\dot{x}_{\overline{H}}> \bullet \nabla\phi + <\dot{x}_S> \bullet \nabla\phi = \frac{d_H\phi}{dt} + \frac{d_S\phi}{dt} \qquad (35)$$

where we name

$$d_H\phi = \lim_{\Delta L_L \gg \lambda_c, \lambda_q} \{\frac{\partial\phi}{\partial t} + <\dot{x}_{\overline{H}}> \bullet \nabla\phi\}\delta t \qquad (36)$$

as "*dynamic differential*" and

$$d_S\phi = \lim_{\Delta L \gg \lambda_c, \lambda_q} [<\dot{x}_S> \bullet \nabla\phi]\delta t \qquad (37)$$

as "*stochastic differential*".

Under the range of validity of equation (46) (i.e., structureless punt-like particles, interacting by L-J central symmetric potential that do not undergo to chemical reactions) the stochastic velocity vector evolves through a pathway that follows the $\phi$-function negative gradient so that

$$\frac{d_S\phi}{\delta t} \text{ is minimum with respect the choice of } <\dot{x}_S> \qquad (38)$$

and $\dfrac{d_S\phi}{\delta t} < 0$ since $<\dot{x}_S>$ is anti-parallel to $\nabla\phi$.

Sometime, some authors speak in term of energy dissipation, so that in this case the criterion (38) reads

$$-\frac{d_S\phi}{\delta t} = |\frac{d_S\phi}{\delta t}| \text{ is maximum with respect the choice of } <\dot{x}_S> \qquad (39)$$

## 5. Stability and maximum stochastic free energy dissipation in quasi-isothermal stationary states

In order to elucidate the significance of the criterion given by (39), we analyze the spatial kinetics far and near equilibrium.

## 5.1 Spatial kinetic equations

By using a well known method [16] we transform the motion equation (34) into a spatial one over a finite volume V. Given a quantity per particle

$$\underline{Y} = \frac{\int_{-\infty}^{+\infty}\int_{-\infty}^{+\infty}\int_{-\infty}^{+\infty} \rho^s Y d^3 p}{\int_{-\infty}^{+\infty}\int_{-\infty}^{+\infty}\int_{-\infty}^{+\infty} \rho^s d^3 p} \tag{40}$$

its spatial density:

$$n\underline{Y} = \int_{-\infty}^{+\infty}\int_{-\infty}^{+\infty}\int_{-\infty}^{+\infty} \rho^s Y d^3 p \tag{41}$$

and its first moment

$$n\underline{Y}\dot{\underline{q}} = \int_{-\infty}^{+\infty}\int_{-\infty}^{+\infty}\int_{-\infty}^{+\infty} \rho^s Y <\dot{q}> d^3 p \tag{42}$$

by using the motion equation (34) it is possible to obtain the spatial differential equation:

$$\partial_t n\underline{Y} + \nabla \bullet n\underline{Y}\dot{\underline{q}} - \int_{-\infty}^{+\infty}\int_{-\infty}^{+\infty}\int_{-\infty}^{+\infty} \rho^s \{\partial_t Y + <\dot{x}_{<H>}> \bullet \nabla Y\} d^3 p$$
$$= \int_{-\infty}^{+\infty}\int_{-\infty}^{+\infty}\int_{-\infty}^{+\infty} Y\{\nabla \bullet \rho^s D\nabla\phi(1 + A + O(\nabla\phi^2))\} d^3 p \tag{43}$$

that by choosing
$$Y = kT\phi, \tag{44}$$
where T is the "mechanical" temperature defined as

$$T = \gamma \frac{<E_{cin} + E_{pot}>}{k} = \gamma\left(\frac{\frac{<p_i><p_i>}{2m} + <\overline{V}_i>}{k}\right), \tag{57}$$

where $\gamma$ is defined at thermodynamic equilibrium, in quasi-isothermal condition and elastic molecular collisions (i.e., absence of chemical reactions) leads to the maximal condition respect to $<\dot{x}_s>$ [14] that reads

$$-\frac{dE_{res}}{dt} = \frac{dTS_{sup}}{dt} = -\iiint_V \left\{ \int_{-\infty}^{+\infty}\int_{-\infty}^{+\infty}\int_{-\infty}^{+\infty} kT\frac{(\phi-1)}{\phi}\rho^s\left(\frac{d_s\phi}{dt}\right) d^3 p \right\} d^3 q = \max \tag{46}$$

where $E_{res}$ is the (free) energy of reservoir that for our purpose can be assumed to work in reversible manner onto the system.
Equation (46) has been proven to be experimentally verified by Sawada in the electro-convective instability [8,17].

## 5.2 Spontaneous free energy increase in far from equilibrium steady-state transition

From general point of view the SQHA system of differential equations is unmanageable as well as the far-from equilibrium kinetics.
In order to gain information about (46) let's analyze it in the simple case of quasi-isothermal stationary states far from equilibrium (chemically and mechanically speaking). This case is still sufficiently general to be interesting since the matter self-assembling phenomena concerning the life generation happen in quasi-isothermal condition.

As a figurative example we take in mind the case of the electro-convective instabilities [17].
In the case of quasi-isothermal system with no chemical reaction taking place, it has been assumed that each infinitesimal volume of fluid is at quasi-thermal equilibrium. On this base, $\Phi$ equals the free energy F, as well as $S^s$ the classical entropy S in equation (46).
In the following we are going to show that the transitions to states with a higher amount of free energy are possible in far from equilibrium kinetics.
To this end, let's consider the overall system with the energetic reservoirs that works reversibly onto the system so that we have

$$\frac{dE_{res}}{dt} = \frac{dF_{res}}{dt} \tag{47}$$

and the heat generated into the system is exchanged with the environment reversibly so that

$$\frac{dF_{sys}}{dt} = 0 \tag{48}$$

then, for the overall system it follows that

$$\frac{dF_{tot}}{dt} = \frac{dF_{res}}{dt} + \frac{dF_{sys}}{dt} = \iiint_V \left\{ \int_{-\infty}^{+\infty}\int_{-\infty}^{+\infty}\int_{-\infty}^{+\infty} kT \frac{(\phi-1)}{\phi} \rho^s \left(\frac{d_s\phi}{dt}\right) d^3p \right\} d^3q < 0 = \text{Max} \tag{49}$$

Moreover, we can also assume that the inequality

$$\frac{dE_{res}}{dt} = \frac{dF_{res}}{dt} < 0 \tag{50}$$

holds for the spontaneous process of energy flowing between the reservoir and the system.
This has been proven in electro-convective instability very far from equilibrium by posing a diode to the electric power in order to prevent that electric energy would flow back into the reservoir. This has been done by Sawada and al. [17] and experiments showed that the current does not spontaneously change sign during transition between instable states.
Let's consider the case that, far from equilibrium in subcritical conditions, the system makes a transition (at time t=0) from a metastable state (1) to another metastable one (2) so that for $t < 0^-$ and $t > 0^+$ we have the sub-system in a stationary state for which it holds

$$\frac{dF_{sys\,1}}{dt} = \frac{dF_{sys\,2}}{dt} = 0 \tag{51}$$

so that by (50) for (t ≠ 0) we have

$$\begin{cases} \frac{dF_{tot}}{dt} < 0 \\ \\ \frac{dF_{sys\,1}}{dt} = \frac{dF_{sys\,2}}{dt} = 0 \qquad i = 1,2 \end{cases} \tag{52}$$

Given (50), in principle, the following cases are possible when the system makes the transition between states 1 and 2. Case "a":

$$\begin{cases} \delta F_{tot\,(1\to 2)} < 0 \quad ; \\ \\ \delta F_{sys\,(1\to 2)} > 0 \quad ; \end{cases} \tag{53}$$

case "b":

$$\begin{cases} \delta F_{tot\,(1\to 2)} < 0 & ; \\ \delta F_{sys\,(1\to 2)} < 0 & ; \end{cases} \tag{54}$$

Hence, given that for the inverse transition 2→1 it would result

Case "a"

$$\begin{cases} \delta F_{tot\,(2\to 1)} < 0 & ; \\ \delta F_{sys\,(2\to 1)} < 0 & ; \end{cases} \tag{55}$$

case "b":

$$\begin{cases} \delta F_{tot\,(2\to 1)} < 0 & ; \\ \delta F_{sys\,(2\to 1)} > 0 & ; \end{cases} \tag{56}$$

since transition between metastable happens in both directions, it follows that one of the two following cases must happen

$$\begin{cases} \delta F_{tot\,(1\to 2)} < 0 & ; \\ \delta F_{sys\,(1\to 2)} > 0 & ; \end{cases} \tag{57}$$

or

$$\begin{cases} \delta F_{tot\,(2\to 1)} < 0 & ; \\ \delta F_{sys\,(2\to 1)} > 0 & ; \end{cases} \tag{58}$$

and since $\delta F_{sys\,(2\to 1)} = -\delta F_{sys\,(1\to 2)}$, considering the whole back and forth transition cycle, at least in one of the two directions, we have that

$$\delta F_{sys} > 0 \tag{59}$$

### 5.3 Entropy decrease in incompressible phases

Generally speaking, the increase of free energy does not also mean that the entropy of the system decreases. This depends by how much is the energy change in the transition.
Nevertheless due to the incompressibility of the fluid phase that makes null energy variations due to the volume change, from the general point of view, in isothermal transitions in fluids (as the electro-convective ones) the free energy increase is almost due by the entropy decrease.
In fact given that

$$\delta F_{sys} \cong \delta E - \delta TS = \delta E_{int} + \delta E_{cin} - \delta TS \tag{60}$$

Since in isothermal transitions in fluids the temperature as well as the fluid density is practically constant, for van der Waals type fluids where $E_{int} = E_{int}(T)$ it follows that $\delta E_{int} \cong 0$ and hence that

$$\delta F_{sys} \cong \delta E_{cin} - \delta TS > 0 \tag{61}$$

Moreover, given that for electro-convective instability in fluids [17] Sawada showed that at the transition between states with increasing numbers of vortexes (that gives a measure of the configurational order of the metastable state) the macroscopic kinetic energy of the molecules does not appreciably increases (or even decreases slightly [17]) we have $\delta E_{cin} \cong 0$ and hence that

$$-\delta F_{sys} \cong T\delta S < 0 \tag{62}$$

From (62) we can see that the ability of a system to make back and forth transitions between metastable states as the consequence both of the presence of fluctuations and the tendency to the maximum stochastic free energy dissipation, allows the spontaneous increase of order in far from equilibrium stationary states in incompressible fluid phase.

Finally it is worth mentioning that the possibility of having matter self-assembling in fluids different from water, in principle, allows the possibility to find organized structures and even living ones in presence of liquid phase such the methane ocean of the Saturn's moon Titan as hypothesized by the NASA [18-19].

## 6. Discussion

Actually, equation (49) has been not directly used to obtain (62). Nevertheless, it enters in the mechanism that allows the system to make transitions between metastable states.

Even if a large fluctuation would displace the system from the first stationary state (bringing the system in the neighborhoods of the second one) this fact will be not enough to attract the system toward the new stationary state.

It is here that the tendency to the maximum stochastic free energy dissipation enters and plays its important role.

As shown by Sawada and Suzuki in electro-convective instabilities [8,17], the tendency to maximize the free energy is the real force that drives the system toward the new stationary condition. In fact, they showed that: 1) The stationary state with the higher free energy dissipation is the more stable (i.e., with a longer living time before the transition to another metastable state); 2) In reaching the fully stationary condition, the free energy dissipation of the system increases and reaches the top at the establishing of final stationary configuration.

Therefore, we can depict the following mechanism: the maximum free energy dissipation generates a basin of attraction [20] (possibly larger, higher is the dissipation rate of the metastable state) for each metastable state and the fluctuations makes the system to jump between them.

## 7. Conclusion

In the present work the SQHA is used as microscopic model to obtain the macro-scale non-equilibrium kinetics for a real gas of L-J interacting particles and its fluid phase.

In the case of particles whose mean distance is bigger than the quantum non-local length that for L-J interaction potential is of order of the molecular interaction distance (so that particles do not have bounded states (e.g., rigid sphere approximation)) is possible to describe the SQHA evolution by means of a Fokker-Plank equation holding even far from the local thermodynamic equilibrium.

In gasses and Marcovian liquids, it is possible to define the kinetic equation of motion as a function of the mean phase space molecular volume that shows maximizing the dissipation of the stochastic part of the SQHA-free energy.

In the case of a real gas with no chemical reactions and at quasi-isothermal conditions, the principle disembogues into the maximum free energy dissipation confirming the experimental outputs of electro-convective instability.

In this case, the model shows that the transition to states with higher free energy can happen and that the increase of free energy in incompressible fluids is almost given by a decrease of entropy leading to the order increase and hence to matter self-organization.

The output of the theory showing that the generation of order, via energy dissipation, is more efficient in fluids than in gasses, because of their incompressibility, leads to the re-conciliation between physics and biology furnishing the eplanation why the life was born in water.

The theoretical output also suggests that the life searching, out of the earth, must consider the possibility to find organized structures in presence of liquid phases different from water.